\documentstyle[aps,prl,psfig,epsfig,floats]{revtex}
\begin{document}
\draft
\twocolumn[\hsize\textwidth\columnwidth\hsize
\csname @twocolumnfalse\endcsname
\title{
Analysis of the elementary excitations in high-$T_c$ cuprates:\\
explanation of the new energy scale observed by ARPES}
\author{D. Manske$^{1}$, I. Eremin$^{1,2}$, and K.H. Bennemann$^1$}
\address{$^1$Institut f\"ur Theoretische Physik, Freie Universit\"at 
Berlin, D-14195 Berlin, Germany}
\address{$^2$Physics Department, Kazan State University, 420008 Kazan, Russia}
\date{April 9, 2001}
\maketitle
\begin{abstract}
  Using the Hubbard Hamiltonian we analyze the energy- and 
  momentum-dependence of
  the elementary excitations in high-T$_c$ superconductors 
  resulting from the coupling to
  spin fluctuations. As a
  result of the energy dependence of the self-energy $\Sigma ({\bf k},
  \omega)$, characteristic features occur in the spectral density
  explaining the 'kink' in recent ARPES experiments. We present
  results for the spectral density $A({\bf k}, \omega)$ resulting from
  the crossover from Im $\Sigma ({\bf k}, \omega)\propto \omega$ to
  Im $\Sigma ({\bf k}, \omega)\propto \omega^2$, for the feedback of
  superconductivity on the excitations, and for the superconducting
  order parameter $\Delta({\bf k}, \omega)$. These results relate also
  to inelastic neutron scattering and tunneling experiments and
  shed important light on the essential ingredients a theory of the
  elementary excitations in the cuprates must contain.
\end{abstract}
]
\narrowtext
For understanding the high-T$_c$ cuprates their elementary excitations
are of central significance.  Angle-resolved photoemission
spectroscopy (ARPES) is a powerful tool for studying the
elementary excitations in high-$T_c$ superconductors because the
spectral density contains all information about self-energy effects.
Due to an improved angular resolution (momentum distribution curve
(MDC) and energy distribution curve (EDC)), data became available which
provide new insight on the momentum and frequency dependence of the
self-energy $\Sigma({\bf k},\omega)$. In particular, a 'kink' feature
at $\hbar \omega \sim $ 50 $\pm$ 15 meV has been observed in 
hole-doped cuprates like Bi$_2$Sr$_2$CaCu$_2$O$_8$, 
Pb-doped Bi$_2$Sr$_2$CuO$_6$, and La$_{2-x}$Sr$_x$CuO$_4$ 
\cite{valla,johnson,kaminski,shen,shen1}. Experiments by Shen 
{\it et al.,}\cite{shen,shen1} observe the kink feature in all directions 
in the first Brillouin Zone (BZ). It exists in both the normal and 
superconducting states. On the other hand, 
Kaminski {\it et al.,}\cite{kaminski} discuss the break only along 
the $(0,0) \rightarrow (\pi,\pi)$ direction occuring when one goes 
from the normal to the superconducting state. Therefore, they did not 
analyze the feature observed by the other group\cite{shen,shen1}.
However, it is quite interesting that a close analysis of data 
of Kaminski {\it et al.}\cite{kaminski} 
in the normal state reveals the same changes 
of the Fermi velocity, $v_{F}$, as noted by 
Shen {\it et al.}\cite{shen,shen1}.  
Thus, there seems to exist a 'new' energy scale
in hole-doped cuprates.
Remarkably, the electron-doped counterparts
({\it e.g.} $\mbox{Ne}_{2-x}\mbox{Ce}_x\mbox{CuO}_4$) do not
show a 'kink'\cite{nagaosa}.
So far, interpretations are given in terms of the presence of a
strong electron-phonon interaction \cite{shen1,nagaosa}, stripe
formation \cite{hanke}, or coupling to a resonating
mode\cite{kaminski,norman}. It is interesting that the experiments 
also observe a change in the dispersion of the elementary excitations 
going from the normal to the superconducting 
state\cite{valla,johnson,kaminski}.
We will show that this results from the feedback effect of 
superconductivity on the elementary excitations.
\begin{figure}[t]
\vspace{-0.5cm}
\centerline{\epsfig{clip=,file=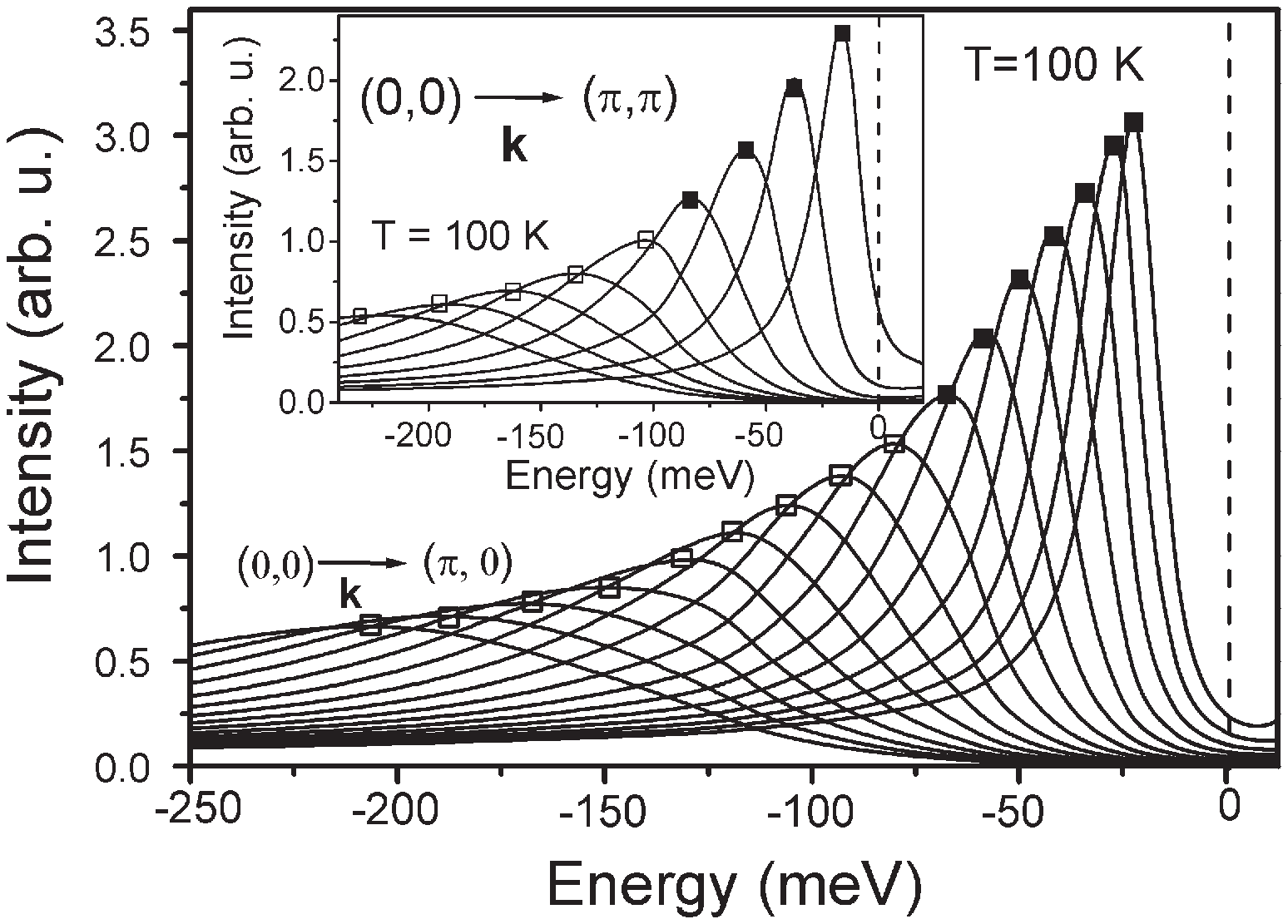,width=7.7cm,angle=0}}
%%%
%\vspace{0.5ex}
\caption{Calculated self-energy effects in the spectral
  density of the quasiparticles in hole-doped superconductors in the
  normal state at $T=100$K and along the $(0,0)\to(\pi,0)$
  direction. The dashed line at $\omega=0$ denotes the unrenormalized
  chemical potential. In the inset the spectral density 
  along the $(0,0)\to(\pi,\pi)$ direction is shown. In both cases 
  at energies about $\hbar\omega \approx 65 \pm 15 $meV a 'kink' 
  occurs, since
  the velocity of the quasiparticles changes.The results are in good
  agreement with experiments (see, for example Fig. 3 in Ref.
  \protect\onlinecite{shen} or Fig.4b in Ref. \protect\onlinecite{shen1}). 
  Note, the width of the spectral density peak is for $(0,0)\to(\pi,0)$ 
  twice the one for $(0,0)\to(\pi,\pi)$.}  
\label{fig1}
\end{figure}

In this letter we present a study of the spectral 
density $A({\bf k},\omega)$ of
the elementary excitations using an electronic theory based on
Cooper-pairing due to an exchange of antiferromagnetic spin
fluctuations.  In particular, we show that the 'kink' in the spectral
density can be naturally explained from the interaction of the
quasiparticles (holes) with spin fluctuations. In agreement with
recent experiments we will demonstrate that the 'kink' feature is
present in both the normal and superconducting
state\cite{johnson,shen,shen1,nagaosa}.  Thus, we are able to explain
recent ARPES experiments which study in detail the spectral density
and in particular the energy dispersion $\omega (\mbox{\bf k}) =
\epsilon({\bf k}) + \Sigma({\bf k},\omega)$.  It is significant 
that the self-energy $\Sigma({\bf k},\omega)$ resulting from
the scattering of the quasiparticles on spin fluctuations can explain
the main features observed. We argue that our results for the elementary
excitations suggest a crossover from Fermi liquid to a non-Fermi
liquid behavior. 
%We also analyze the feedback effects due to
%superconductivity on the elementary excitations and find fair
%agreement with Ref.  \onlinecite{kaminski}. 
Furthermore, we obtain a
picture consistent with inelastic neutron scattering (INS) and
tunneling measurements\cite{manske}.

The theoretical analysis is based on the spectral density
for the elementary excitations\cite{appendix}   
within Nambu space\cite{schrieffer} which are given by, after 
continuation to the real $\omega$-axis, 
\begin{equation}
A({\bf k},\omega) =
-\frac{1}{\pi}\,\frac{\Sigma''({\bf k},\omega)}%
{\left[\omega - \epsilon_{\bf k} - \Sigma'({\bf k},\omega)
\right]^2 + \left[\Sigma''({\bf k},\omega)\right]^2}
\quad .
\label{spectral}
\end{equation}
\begin{figure}[t]
\vspace{-0.1cm}
\centerline{\epsfig{clip=,file=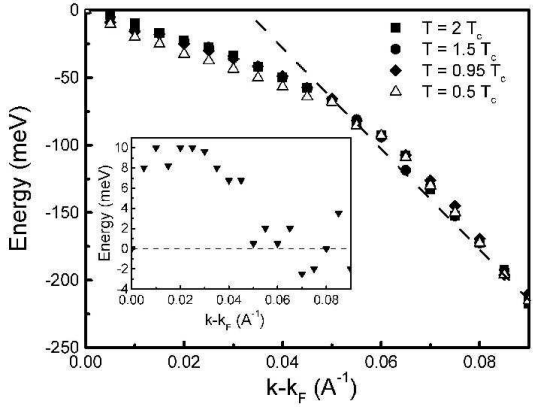,width=7.7cm,angle=0}}
\vspace{0.5ex}
\caption{Positions of the peaks in the spectral density $A({\bf k},\omega)$ 
  versus {\bf k-k}$_F$ (energy dispersion)  
  along the $(0,0)\to(\pi,0)$ direction of the BZ calculated within the 
  FLEX approximation. 
  This has to be compared with the position of the peaks derived 
  from the momentum distribution curve (MDC) for a hole-doped
  superconductor as measured in experiment. 
  The curves show a 'kink' at energies about 
  $\hbar\omega\approx 65 \pm 15$meV.  
  The dashed line is a guide to the eyes. We find small
  changes due to superconductivity which almost coincide with the
  'kink' position. Inset: change in the peak position in 
  A({\bf k}, $\omega$) in the superconducting state ($T=0.5T_c$). 
  The results are in fair agreement with ARPES data
  \protect\onlinecite{kaminski}.}
\label{fig2}
\end{figure}
Here, $\epsilon_{\bf k}$ is a tight-binding energy dispersion on a
square lattice and $\Sigma'({\bf k},\omega)$ and $\Sigma''({\bf
  k},\omega)$ are the real and imaginary part of the self-energy,
respectively.  We perform our calculations for the elementary excitations
\begin{equation}
\omega({\bf k}, T) = \epsilon({\bf k}) + \Sigma({\bf k},\omega({\bf k},T), T)
\quad .
\label{eq:6}
\end{equation}
The superconducting gap function $\phi({\bf k},\omega)$
is calculated self-consistently using
the 2D one-band Hubbard Hamiltonian for a $\mbox{CuO}_2$-plane,
which reads on a square lattice
\begin{equation}
H = - \sum_{\langle ij \rangle \, \sigma}
t_{ij}\left( c_{i\sigma}^+ c_{j\sigma} +
c_{j\sigma}^+ c_{i\sigma}\right)
+ U\, \sum_i n_{i\uparrow}n_{i\downarrow}
\quad .
\label{eq:hubbard}
\end{equation}
Here, $c_{i\sigma}^+$ creates an electron with spin $\sigma$
on site $i$, $U$ denotes the on-site Coulomb interaction,
and $t_{ij}$ is the hopping integral. The imaginary part of the 
self-energy is given by\cite{manske}
\begin{eqnarray}
{\small
\mbox{Im } \Sigma({\bf k}, \omega)} & = & {\small -\frac{U^2}{4} 
\int d \omega' \left[ \coth \left(
\frac{\omega'}{2T}\right) - \tanh \left(
\frac{\omega'-\omega}{2T}\right) \right]} \nonumber\\
& \times & {\small \sum_{\bf k'} \mbox{Im} \, 
\chi ({\bf k-k'}, \omega') \,\, \delta(|\omega - \omega'| - \epsilon_{\bf k'})}
\label{eq:sigma}
\end{eqnarray}
where Im $\chi ({\bf q}, \omega)$ is 
the imaginary part of the spin susceptibility within the random phase 
approximation. We determine the coupling of the quasi-particles to the 
spin fluctuations using an effective perturbation theory 
(FLEX)\cite{bickers,dahm,moria} which we calculate directly on the real 
$\omega$-axis.

These equations are standard, however it is
important to realize that due to the combined effects of Fermi surface
topology and $\chi({\bf q}={\bf Q},\omega)$ at the antiferromagnetic
wave-vector {\bf Q}$_{AF}$ = ($\pi, \pi$), the {\bf k} and $\omega$
dependence of $\Sigma ({\bf k},\omega)$ become very pronounced and
change the dispersion $\omega({\bf k})$. It is known that the strong
scattering of quasiparticles on antiferromagnetic spin fluctuations
results in a non-Fermi liquid behavior of the quasiparticle self-energy
for low-lying energy excitations, in particular, in Im $\Sigma \sim
\omega$\cite{varma,ruvalds}.  Clearly, it follows already from the Eq.
(\ref{eq:6}) that the expected doping and momentum dependence
resulting from the crossover from $\Sigma \propto \omega^2 $ to
$\Sigma \propto \omega $, {\it i.e.} to a non-Fermi liquid behavior,
can be reflected in $\omega({\bf k})$ and $A({\bf k}, \omega)$.
Simply speaking, the change in the $\omega$-dependence of 
the self-energy $\Sigma
({\bf k},\omega)$ changes the velocity of the elementary excitations.
Thus, for a given ${\bf k}$-vector, the MDC curve shows a 'kink' at
some characteristic frequency controlled by $\omega_{sf}$ 
($\omega_{sf}$=spin fluctuation energy). Regarding the superconducting state
the {\bf k}- and $\omega$-dependence of the order parameter
$\Delta({\bf k}, \omega)$ is important and yields the feedback of the
superconducting state on the elementary excitations.

This quite new structure in $\omega({\bf k}, T)$ which
is present in both the normal and superconducting state is shown in
the figures exemplarily for optimal doping 
and results from our calculations obtained by
solving the above equations self-consistently within
a conserving approximation\cite{dahm}. 
The full momentum and frequency dependence
of the quantities is kept and no further parameter is introduced.

In Fig. \ref{fig1} we present results for the frequency and momentum
dependence of the spectral density in the normal state exemplarily 
along the $(0,0) \rightarrow (\pi,0)$ direction calculated 
using the canonical parameters $U = 4t$, and $t$ = 250 meV\cite{refereeD}. 
The changes in
the {\bf k}-dependence of the peak in $A({\bf k}, \omega)$ reflect
the characteristic features in the self-energy $\Sigma({\bf k},
\omega)$ or in the velocity $v_{\bf k}$ of the quasiparticles. The
kink occurs at energies about 
$\hbar \omega \approx 65 \pm 15$ meV for optimal doping (x = 0.15)
and T$_c \approx $ 65K. 
We also find that the 'kink' feature 
is present in all directions in the BZ ($\omega \approx 
\omega_{sf}+ {\bf v}_F (\phi) {\bf k}$, {\bf k=k}($k,\phi$)) 
and, in particular, 
along the diagonal $(0,0)\rightarrow (\pi, \pi)$ direction as shown in the 
inset of Fig.1.  We get that the kink is similar pronounced in both 
directions. Moreover, we see
from our calculations that this feature has only a weak
temperature dependence over a wide temperature range. It changes
only at very small temperatures which we will describe later.

In Fig. 2 we show the positions of the peaks along $(0,0)\to (\pi,0)$ 
shown in Fig. 1 as a function of 
(${\bf k} - {\bf k}_F$) for different temperatures.  We
obtain only small changes due to superconductivity which almost
coincide with the 'kink' position. Remarkably, the deviation at ${\bf
  k} - {\bf k}_F \approx $ 0.05 A$^{-1}$ is due to the frequency
dependence of the self-energy and reflects the transition from
Fermi-liquid to a non-Fermi liquid behavior along the route ($0, 0$)
$\rightarrow$ ($\pi, 0$) in the Brillouin Zone (BZ) for both normal
and superconducting state. In the inset we show results for the
difference in the peak positions for the normal and superconducting
state along $(0,0) \rightarrow (\pi,0)$ direction 
in order to see also the feedback of superconductivity. Note,
this disappears for ${\bf k} - {\bf k}_F \approx $ 0.05 corresponding
to approximately 65meV. We get changes of about 10 meV while  
Kaminski {\it et al.,}\cite{kaminski} observe along 
$(0,0)\rightarrow(\pi,\pi)$ direction a larger difference of 
about 20 meV. By inspecting Eq. (\ref{eq:sigma}) this is expected, 
since the feedback effect of superconductivity on $\chi$ is larger
for {\bf Q}$\approx(\pi,\pi)$.
%In order to understand the feedback of
%superconductivity and crossover from non-Fermi to a Fermi liquid
%behavior we analyze the frequency and temperature dependence of the
%self-energy more in detail.

In order to investigate the effect of the self-energy $\Sigma ({\bf
  k}, \omega)$ on the dispersion $\omega({\bf k}, T)$ 
we show in Fig. 3 results of our calculations for Im
$\Sigma({\bf k}_n, \omega)$ at the wave vector along the node line of
the superconducting order parameter in the first BZ. The transition
from $\Sigma({\bf k}, \omega) \propto \omega^2$ to $\Sigma({\bf k},
\omega) \propto \omega$ for low-lying frequencies is shown for various
temperatures. Note, the deviation
\begin{figure}[t]
\vspace{-0.3cm}
\centerline{\epsfig{clip=,file=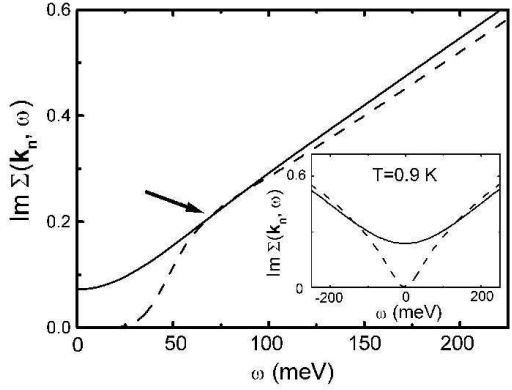,width=7.5cm,angle=0}}
\vspace{0.5ex}
\caption{Calculated frequency dependence of the quasiparticle
  self-energy at the node $\Sigma({\bf k}_n,\omega)$, ${\bf
    k_n}=(0.4,0.4)\pi$.  The solid curves correspond to the normal
  state at $T=2T_c$, whereas the dashed curves refer to the
  superconducting state at $T=0.5T_c$.  At ${\bf k}={\bf k_n}$, where
  the superconducting gap vanishes, one clearly sees
  approximately  at $\hbar\omega=65$meV a crossover from Fermi liquid
  behavior ($\Sigma \propto \omega^2$) to a non-Fermi liquid behavior
  ($\Sigma \propto \omega$) for low-energy frequencies as a function
  of temperature. We show in the inset the
  behavior of $\Sigma({\bf k}_n,\omega)$ calculated at very low
  temperatures $T=0.003t\simeq 0.9$K (dashed line).  }
\label{fig3}
\end{figure}
from Landau's theory (see solid curve in Fig. 3), Im $\Sigma \sim
\omega$, results in our picture from the strong scattering of the
quasi-particles on the spin fluctuations and is expected to disappear
at temperatures T$\rightarrow$0, see inset of Fig. 3.  
In particular, the changes in the velocity of quasiparticles are 
determined in EDC as $v_{F}^{*}=v_{F}/
\lgroup 1+\frac{d\Sigma^{'}_{k} (\omega)}{d\omega}\rgroup$ versus frequency. 
At the frequencies around 65 meV the Re $\Sigma_k (\omega)$ shows 
a flattening as can be seen via a Kramers-Kronig analysis of Im $\Sigma$.
Therefore, at this frequency the effect of the scattering on spin 
fluctuations almost disappears. Thus, we find a Fermi liquid behavior. 
Our results also agree with previous ones obtained 
within the spin-fermion model\cite{chubuk}. 
In our microscopic theory we also recover Fermi liquid 
behavior for T$\sim \omega
<< \omega_{sf}$. Here, $\omega_{sf}$ is the characteristic spin
fluctuation energy measured in INS (roughly the peak position of
$\mbox{Im }\chi({\bf Q},\omega)$\cite{euro})
and is typically around $25$meV for
hole-doped superconductors\cite{bapines}. Previously we have 
shown that our $\omega_{sf}$ gives a good 
description of INS data\cite{manske}. 
On the other hand, for
T$<$T$_c$ the scattering is also strongly reduced not only due to
$\omega < \omega_{sf}$, but also due to a feedback effect of
superconductivity which will be discussed in connection with Fig.
\ref{fig4}.

There is a wide discussion whether or not layered cuprate
superconductors behave like conventional Fermi liquids.  Earlier
experiments (for a review, see Ref. \cite{timusk}) reveal non-Fermi
liquid properties, in particular a linear resistivity $\rho(T)$ for
optimal doping, non well-defined quasiparticle peaks above the
superconducting transition temperature T$_c$ seen in
ARPES\cite{campuzano}, and a strong temperature dependence of the
uniform spin susceptibility observed by nuclear magnetic resonance
(NMR) \cite{slichter}.  The
\begin{figure}[t]
\vspace{-0.3cm}
\centerline{\epsfig{clip=,file=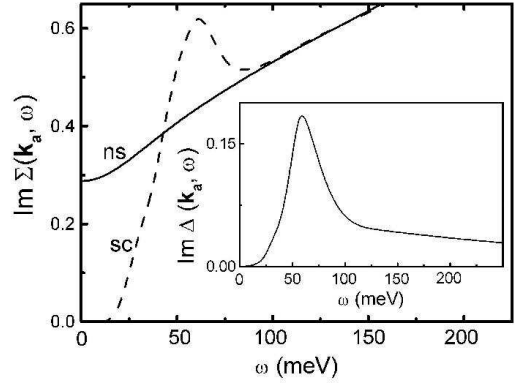,width=7.5cm,angle=0}}
\vspace{0.5ex}
\caption{Calculated frequency dependence of the quasiparticle
  self-energy $\Sigma({\bf k}_a,\omega)$ at the wavevector ${\bf
    k}={\bf k_a}\approx(1,0.1)\pi$ (antinode).  The solid curves
  correspond to the normal state at $T=2T_c$, whereas the dashed
  curves refer to the superconducting state at $T=0.5T_c$. For the
  wavevector ${\bf k_a}$ the feedback effect of superconductivity on
  the self-energy is shown.  Inset: Superconducting gap function
  $\Delta(\omega)$ at wave vector ${\bf k}={\bf k_a}$ versus
  frequency. Since the behavior of $\Delta$ and $\Sigma$  is
  controlled by Im $\chi ({\bf q}, \omega)$ we are able to 
  connect these results to the resonance peak observed
  by INS in cuprates\protect\cite{manske}.}
\label{fig4}
\end{figure}
phenomenological concepts of a Marginal Fermi liquid (MFL) and a
Nested Fermi liquid (NFL) have been introduced in order to explain the
deviations in the normal state from Fermi liquid theory
\cite{varma,ruvalds}.  Our results shed more light on this question.
In agreement with the picture of Ruvalds and co-workers we obtain the
$\omega$- and $T$-dependence of the self-energy mainly due to
scattering of the quasiparticles on spin fluctuations which is
strongest for a nested Fermi topology. This also provides a
microscopic justification for the MFL approach\cite{refereea}.  
Thus, for optimal
doping ($x=0.15$), the microscopic FLEX approximation includes the
phenomenological concepts of both NFL and MFL\cite{mabe}.  It would
be interesting to extend our studies to the underdoped regime,
however, the origin of the pseudogap is still unknown\cite{referE}.

In Fig. 4 we demonstrate the feedback of superconductivity on
$\Sigma({\bf k}, \omega)$. We expect that it is the strongest for {\bf
  k}$\approx$ ($\pi$,0.1$\pi$) where the gap $\Delta(\omega)$ is
maximal. One sees that mainly the superconducting properties in
$\Delta({\bf k}, \omega)$ and in particular in Im $\Delta({\bf k},
\omega)$ induce changes in the self-energy. For the comparison with
the experiment we also present our results for the superconducting
gap. Note, this behavior of $\Sigma({\bf k}, \omega)$ and $\Delta({\bf
  k}, \omega)$ is related also to INS and optical conductivity
experiments. In particular, the peak position of Im $\Sigma({\bf k}_a,
\omega)$ is approximately at $3\Delta_{0} - \omega_{sf} \approx 
\omega_{res} + \Delta_{0}$ ($\omega_{res}$ denotes the resonant 
frequency observed in INS) according to our
previous analysis\cite{manske}. This is in a good agreement 
with results obtained within the frame of the 
spin-fermion model\cite{chubuk}. 
%It is important that $\Delta(\omega)$
%also agrees well with the tunneling experiments\cite{manske}.

It is remarkable that for electron-doped superconductors with a
different dispersion $\epsilon_k$\cite{prb2000}, in particular with a 
flat band lying 300meV below $\epsilon_F$ at ($\pi, 0$), 
we get no 'kink' feature up to frequencies about 100meV.  
This is also in agreement with experiment\cite{nagaosa}.  
The reason behind this is that Im $\chi({\bf q},
\omega)$ has a peak at larger frequencies and which is much less
pronounced than for hole-doped cuprates\cite{euro}.

In summary, calculating the pronounced momentum and frequency
dependence of the quasiparticle self-energy $\Sigma$ in hole-doped
high-T$_c$ cuprates we find that this results in a 'kink' structure in
the dispersion $\omega({\bf k})$ which agrees well with recent ARPES
experiments.  For describing the physics in the cuprates it is
important that the origin of this is the coupling of the
quasiparticles to the spin fluctuations.  The reason for the kink
structure is a change in the $\omega$-dependence of the self-energy
$\Sigma$ from non-Fermi liquid to a Fermi liquid behavior.  Due to a
different spectrum $\mbox{Im }\chi({\bf q},\omega)$ of the spin
fluctuations in electron-doped cuprates we do not find a 'kink' in the
corresponding spectral density.  Furthermore, the feedback effects due
to superconductivity on the elementary excitations clearly reflect the
symmetry of the superconducting order parameter. The calculated
density of states $N(\omega)\equiv A(\omega) = 1/ N \sum_{\bf k}
A({\bf k}\omega)$ compares well with SIN tunneling data\cite{manske}. 
However, due to spatial averaging such experiments do not 
exhibit a kink structure.

Its pleasure to thank J. Mesot, M.~S. Golden, S. Borisenko, 
D. Fay, R. Tarento, and P. Pfeuty for useful discussions. 
We are grateful to German-Franch foundation (PROCOPE) for the
financial support. The work of I. E. is supported by 'Alexander von
Humboldt' foundation.

\begin{references}
\bibitem{appendix} The standard formalism yields 
$G = [i\omega_nZ + \epsilon_k + 
\xi]/[\left(i\omega_nZ\right)^2 -
\left(\epsilon_k + \xi\right)^2-\phi^2]$ with the self-energies 
$\Sigma^G = i\omega_n\left[ 1-Z \right]
+\xi $, and $\Sigma^F = \phi$
where the latter is the anomalous part.  In the
normal part of the self-energy, $\Sigma^G$, $Z$
denotes the mass renormalization and $\xi$ describes the
renormalization of the tight-binding dispersion $\epsilon_k$.
\bibitem{refereeD} Our calculations have been performed in the 
"clean-limit" and thus do not consider the additional
scattering of the carriers on impurities. Also the intrinsic width of the 
spectral function due to scattering on the antiferromagnetic 
spin fluctuations provides a small uncertainty.
\bibitem{refereea} Note, although the results of our microscopic 
calculations agrees with MFL theory, we do not 
prove the existence of a quantum critical point suggested there.
\end{references}
\end{document}